# ELECTRONICS AND PHOTONICS:
# TWO SCIENCES IN THE BENEFIT OF SOLAR ENERGY CONVERSION

Mihaela Girtan

Photonics Laboratory, Angers University, France

mihaela.girtan@univ-angers.fr

**Abstract**

This paper gives a personal global point of view on two sciences: Electronics and Photonics towards plasmonics and solar energy conversion. The new research directions in these two sciences are pointed out by comparison and the perspective for future new solar devices is presented.

Building with sand…. One of the most abundant chemical compounds in nature is the silicon dioxide, also known as silica, most commonly found in nature in sand or quartz. Starting from sand, the human being intelligence transformed then the silica into silicon and finally quite the whole world today based on electronics and photonics, with computers, optical transmission data, mobile phones, solar cells, use sand as starting material. Of course there are other elements and sand alone couldn't be use to get all these ends.

Beside inorganic materials, the new research achievements show that organic materials could be also used successfully for electronics and photonics. The new trends in physics and technology are the plastic electronics and plastic photonics.

Most recent studies proved that in the next future the sand will be replaced by graphene.

This paper presents a parallel and the equivalence between the electronics and photonics. Starting from electron in electronics, photon, solitons and plasmons in photonics, electrical cables – optical fibers, plasmonic wave guides, electrical circuits - optical circuits, electrical transistors – optical transistors, plasmonster, electrical generators – pulsed lasers and spasers, photonics gets step by step all the tools already existing in electronics.

Solar energy could be converted in many ways, the most known is the conversion in electricity. Today we need that the energy is in form of electricity because most of the apparatus that we use are based on electricity: informatics, motors, etc.

However the progress in photonics with optical circuits, optical transistors, etc., shows that the photonics informatics could be possible. Also the optical manipulation and optical engines concept were already demonstrated experimentally.

If the laser propulsion will be achieved, and the optical engines will work, the question that will rise tomorrow is "Shall we still use the electricity in the future? What will be the solar devices tomorrow ?".

## 1. Introduction

The continuously rising demand on energy is one of the most important problems today. The 2009 World Energy Outlook, published by the International Energy Agency, predicts that world demand for oil (often used as reference for world demand for energy) will increase from 2,000 million tons of oil equivalent (mtoe) to 16,800 mtoe in 2030. The increase in worldwide energy demand is principally driving by the increase of population, the industrialisation and the globalization. Analysing the final consumption energy by sector we remark, during the last 40 years, an equal repartition (about 30%) between industry, transport and residential.

The fluctuations of oil prices during the last years, due to the volatility of the financial markets and economic turmoil, have highlighted our strong dependence on oil and have added an additional argument for using and developing other energy resources.

The most important sustainable, renewable green energy source is the Sun which is delivering in an hour more energy ($4.3 \times 10^{20}$ J) to Earth than the energy we use in one year using fossil, nuclear and other renewable sources together ($4.1 \times 10^{20}$ J). Beside the other renewable energies sources: biomass, wind, hydroelectricity, the solar energy source has many advantages: is ready available, secure from geopolitical tension, available everywhere, even on isolated sites and less polluting.

Solar energy is world-wide abundant and photovoltaic is at present a key technology option to realise the shift to a decarbonised energy supply. In 2009, the photovoltaic industry production increased by more than 50% and reached a world-wide production volume of 11.5 GWp of photovoltaic modules. Yearly growth rates over the last decade were in average more than 40%, which makes photovoltaics one of the fastest growing industries at present. However, the development of photovoltaic market is still slow down by the high fabrication cost of the solar cells panels. Looking just on different chart concerning the repartition of energy in function of resources for different countries we can see that the place of renewable energies is still very small (geothermal/solar/wind represents all together only 0.8 % in the whole world).

Most of these future renewable energy solutions are bases on the conversion of another form of energy (mechanic, thermal, electromagnetic) in electricity for different uses (informatics, motors, lighting, heating) which mean to transform the electrical energy in other forms of energies again. But, will "electricity" remains in the future the preferred form to exploit and use the energy? What will be the solar energy tomorrow devices?

This paper gives a global point of view on two sciences: Electronics and Photonics towards plasmonics and solar energy conversion. The new research directions in these two sciences, based on nanotechnologies, are pointed out and the perspectives for future new solar devices without transforming the solar energy in electricity are presented.

## 2. Experimental and discussions

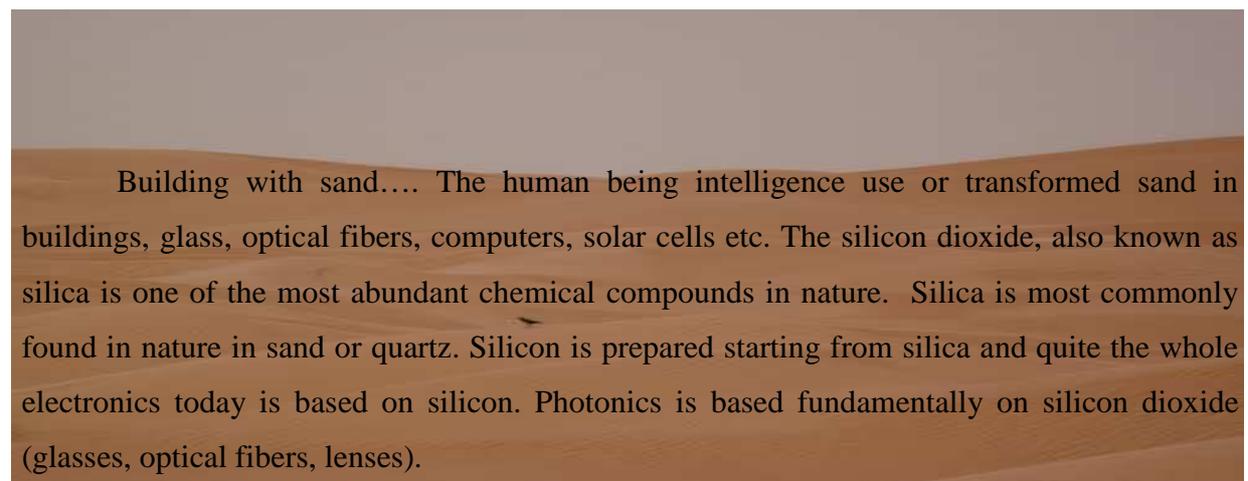

Building with sand…. The human being intelligence use or transformed sand in buildings, glass, optical fibers, computers, solar cells etc. The silicon dioxide, also known as silica is one of the most abundant chemical compounds in nature. Silica is most commonly found in nature in sand or quartz. Silicon is prepared starting from silica and quite the whole electronics today is based on silicon. Photonics is based fundamentally on silicon dioxide (glasses, optical fibers, lenses).

So finally the whole world today, based on electronics and photonics, with computers, optical transmission data, mobile phones, solar cells, use sand as starting material (Fig.1). Of course there are other elements and sand alone couldn't be use to get all these ends.

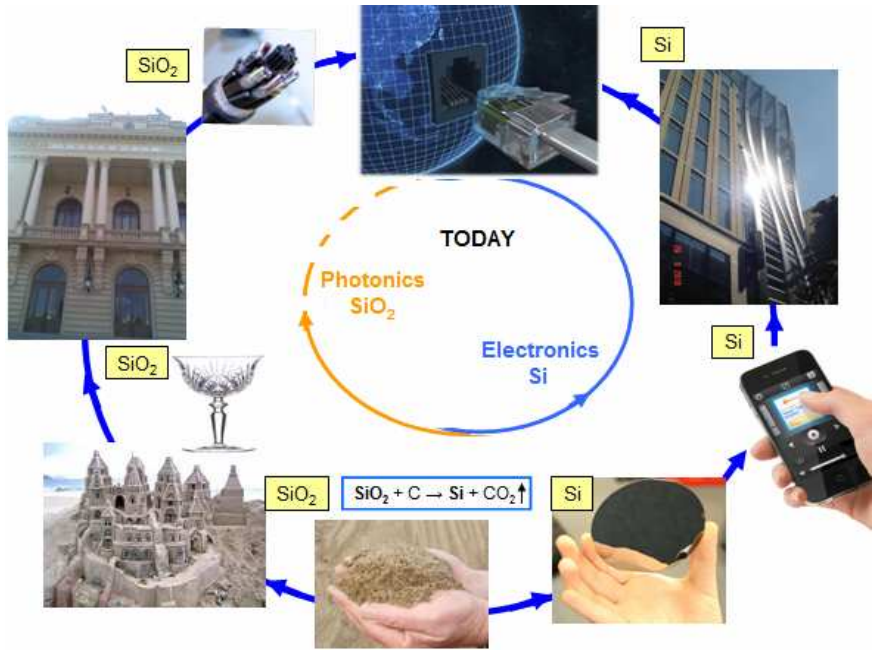

Fig.1. Sand as material source in our day life: for buildings, glasses, optical fibers, electronics, solar cells…

The progress of nanotechnology conduct to a fast development of these two sciences: electronics and photonics. First were the development of electronics and, more and more, today the nanotechnologies are present in photonics research. Much more, if we compare the basic elements of these two sciences: cables, generators, information carriers, etc., we remark that there are a lot of elements in common (Fig.2).

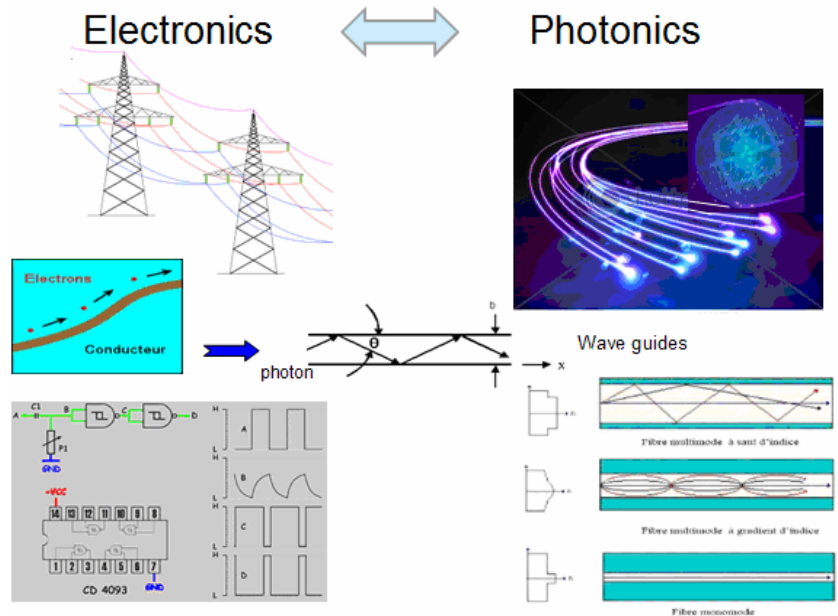

Fig.2. Information carrier vectors and the transports ways of the carriers in electronics and photonics

Today computers and informatics are based on electron and electronics circuits. But will be possible to do with photons or other carrier vectors what we do today with electrons? In place of introducing electrons in computers or mobile phones circuits, will be possible to introduce photons? Can we imagine that tomorrow informatics will be based on photonics, or plasmonics circuits?

Four step to pass from electronics informatics to photonics informatics (Fig.3)…

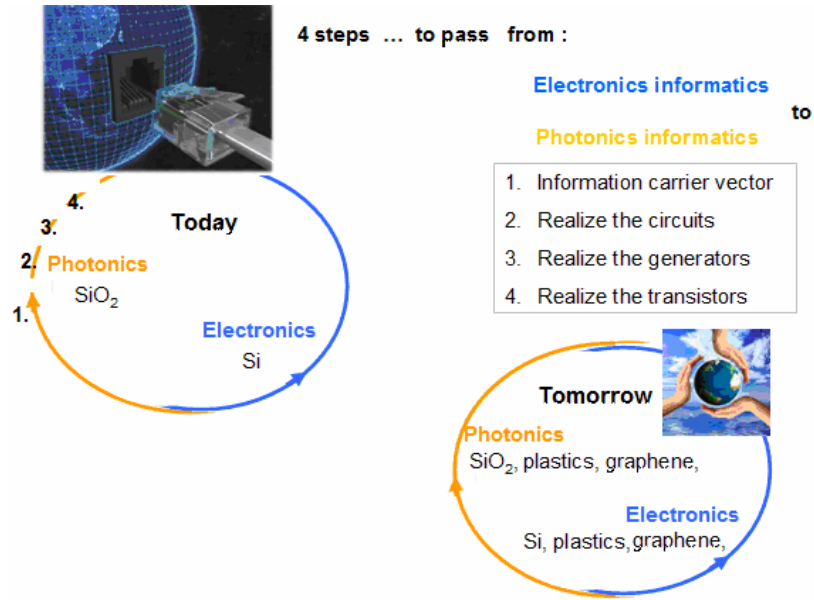

Fig.3. Today and tomorrow electronics and photonics: four steps to pass to photonic computers

To switch from "electronic computers" to "photonics computers", we should found in photonics the equivalents at least of the most important elements existing in electronics: 1) the information carrier vectors 2) the cables an circuits 3) the generators, 4) the transistors.

**First step – choosing the equivalent information carrier vector**

For "photonics computers", beside photons others information carrier's vectors could be solitons, light balls, or plasmons. Plasmon is a quasi-particle associated to plasma oscillations of free electrons density (jelium model) in respect with positive ions positions (Fig.4.) [1].

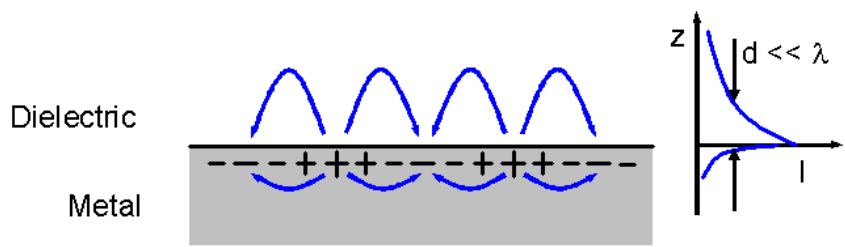

Fig.4. Plasmon propagation along a metal/dielectric interface

This particle associating the existing electrons presenting in the material and the injected photons, opens the Era of nanophotonics. Plasmons offer some very important unique advantages such as: to transmitting the information with higher frequencies (~100 THz) than those employed today in data transmission or to confining light in very small dimensions objects.

**Second step – constructing the equivalent information carrier transport cables and circuits**

The next step towards photonics informatics is to realize the information carrier's cable and circuits. And here again the parallel between electronics and photonics put in evidence a lot of similarities, and the electronics devices served as model for photonics devices.

Beside inorganic materials, the new research achievements show that organic materials could be also used successfully for electronics and photonics. The new trends in physics and technology are the plastic electronics and plastic photonics.

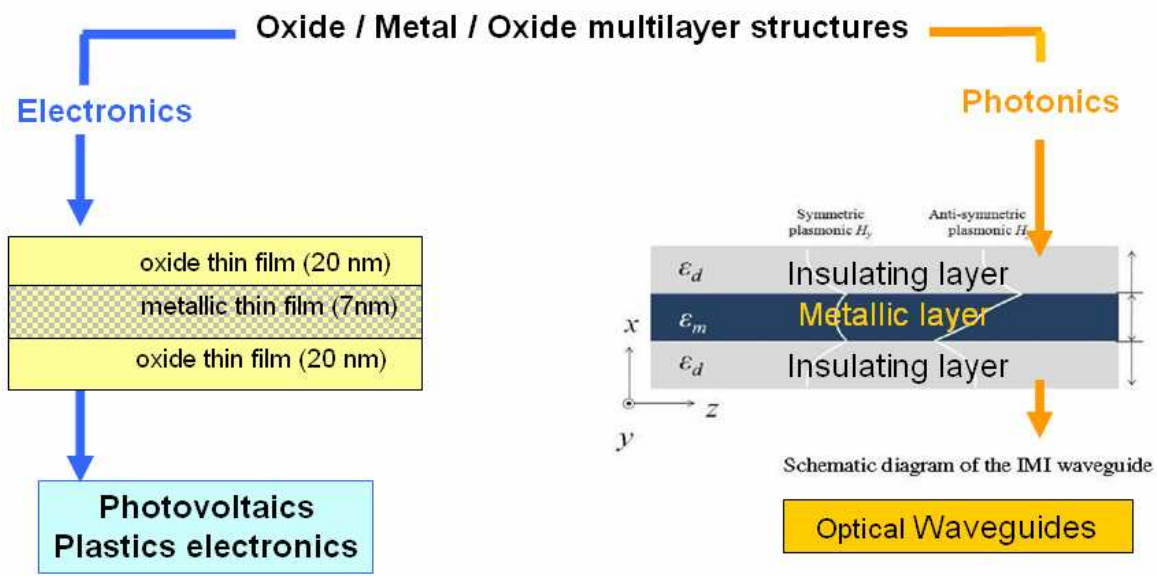

Fig.5. Comparison of multilayer structures used at present in new generation devices in electronics [2] and plasmonic wave guides [3]

In transparent flexible electronics and in third generation solar cells new promising electrodes are composed of three layer oxide/metal/oxide [2]. In photonics we find the same structures for plasmonic wave guides (Fig.5.) [3].

And the comparison could be extended for organic optical wave guides [4] whose architecture is, with small difference, the architecture of an organic solar cell (Fig.6.) [5].

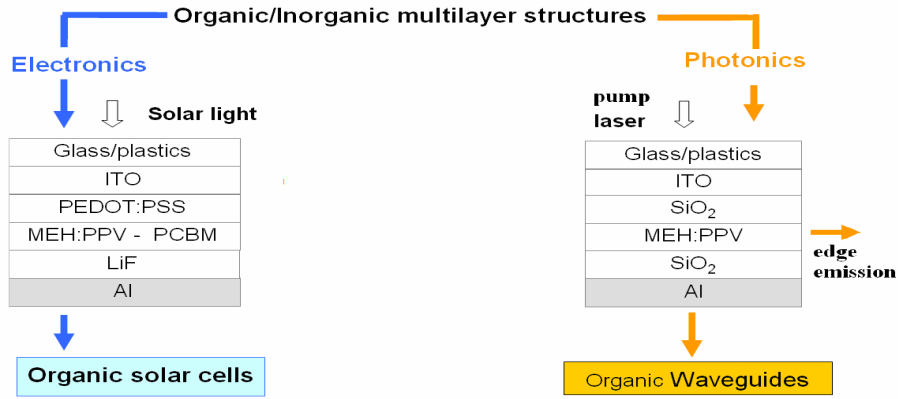

Fig.6. Comparison between organic/inorganic multilayer structures in third generation organic solar cells [5] and organic photonics wave guides [4]

The electronics inspired the photonics for optical circuits and by combining these two science, plasmonics circuits were realized in the last few years (Fig.7.) [6,7].

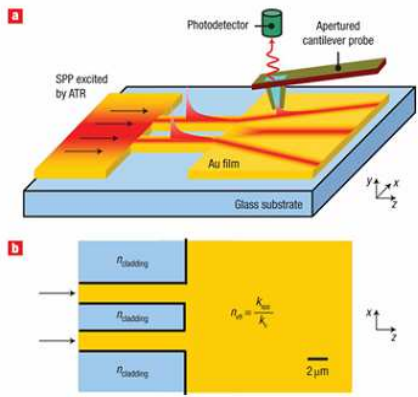
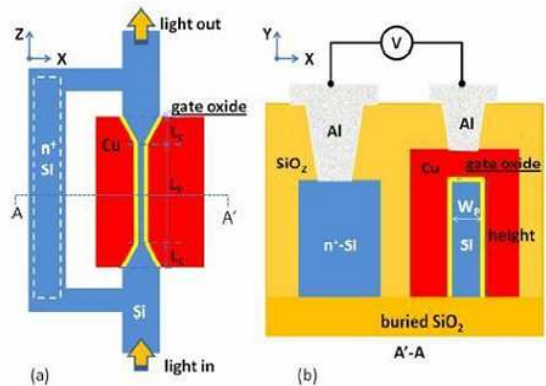

Fig.7. Nanoplasmonics circuits in analogy with microelectronics circuits

**Third step – founding the equivalent for electrical generators**

So in the toolbox of photonics informatics we could find the particles, the cables, the circuits. Let now have a look on generators. Lasers could be seen in photonics devices as the equivalents of electrical generators ("photonic generators").

If since now, the whole electronics and photonics were based on sand (silicon and silicon dioxide), the carbon, in its "graphene" form, might replace in the future the silicon. The graphene, was obtained for the first time in 2004 by two researches of Manchester

University, Andre Geim and Konstantin Novoselov (nobel prize in 2010) by micro-mechanical alleviation of graphite. As material for electronics, graphene is very interesting for plastic applications as a transparent electrode with very good mechanical properties. Much more, the new transfer techniques allow the deposition on large area flexible surfaces [8].

Due to it's very good mechanical and electrical conduction properties the research community were centered since now on its electronics properties, and graphene's optical response has previously been considered to be weak and uninteresting. Different from conventional semiconductors materials, the valence band and conduction band in graphene are smooth-sided cones that meet at a point – called the Dirac point and graphene energy band gap is equal to zero (Fig.8.). Due to this absence of the optical band gap, graphene absorb all photons at any wavelength. However, if incident light intensity becomes strong enough, due to the Pauli blocking principle, the generated carriers fill the valence bands, preventing further excitation of electrons at valance band. Hence this property could be potentially exploited to realize "saturable absorbers" with wide optical response ranging from ultra-violet, visible, infrared to terahertz. By introducing a saturable absorber into the laser cavity, modes with stronger intensity will be selected while modes with weaker intensity will be suppressed, giving rise to short and very intense light pulses [9].

Much more graphene structure specificity and charge transport open new research directions on nanoplasmonics: "graphene nanoplasmonics" [10].

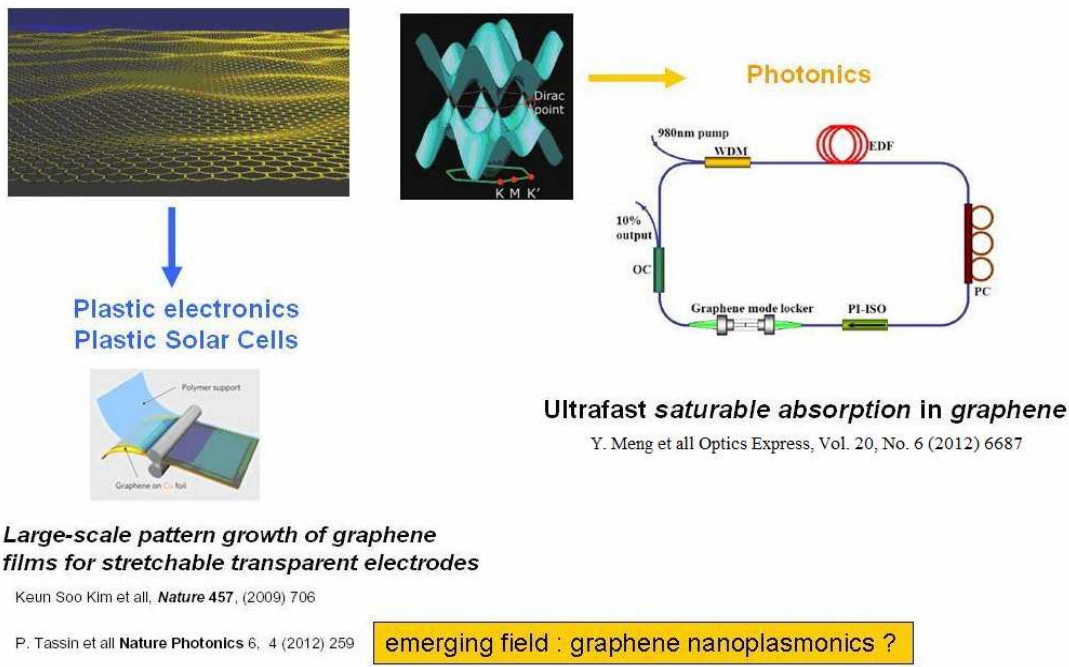

Fig.8. Graphene applications in plastic electronics and fiber lasers.

**Fourth step – founding in photonics the equivalent of electronic transistors**

The last element of the photonic informatics toolbox is the transistor and here again the electronics inspired the photonics. In Fig.9. we present by comparison an organic electronic transistor and the first optical transistor, which is based equally on an organic molecule having similar structure to pentacene molecule generally used in organic transistors [11].

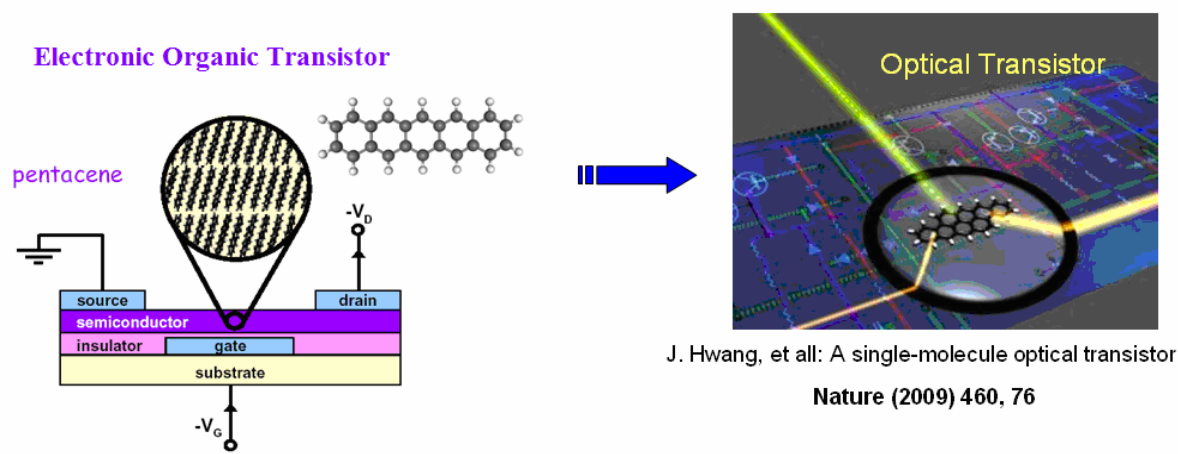

Fig.9. Comparison between an electronic organic transistor and an optical transistor

Starting from electron in electronics, photon, solitons and plasmons in photonics, electrical cables – optical fibers, plasmonic wave guides, electrical circuits - optical circuits, electrical transistors – optical transistors, plasmonster [12], electrical generators – pulsed lasers and spacers [13], photonics gets step by step all the tools already existing in electronics.

3.  **Conclusions**

If the photon informatics will be achieved in the future, that means that it will be possible to exploit directly the solar energy by injecting the photons in the optical (or plasmonics) circuits of tomorrow devices which will be the photonic equivalent of our today electronic devices (mobile phones, computers, etc).

And the question that will rise is: "Did we still need electricity in the future? How will look like tomorrow solar devices?"

What are our principals needs? We need electricity for: lighting, heating, communications, informatics, motors …

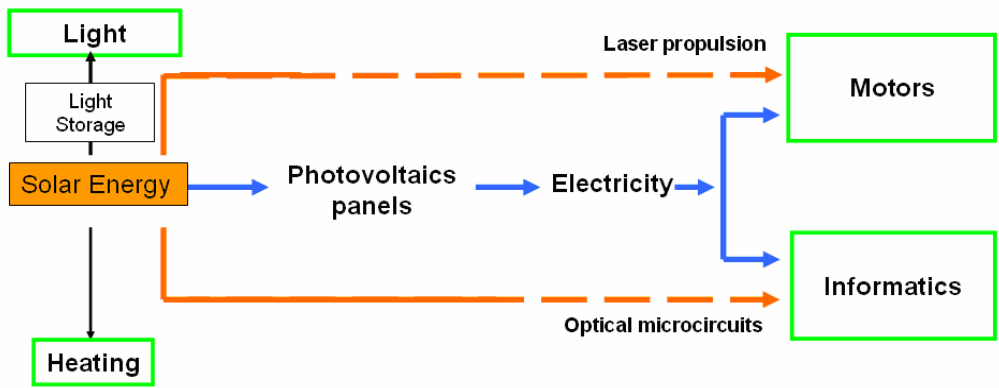
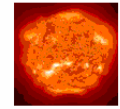

Fig. 10. Tomorrow possible exploitation of solar energy without it's transformation in electricity

Today we transform different forms of energy into electricity to achieve most of these needs (Fig.10). But, it will be possible in the future to avoid the transformation of the energy in electricity and exploit directly the solar energy for all our today life requirements? For heating, we can use and we already use, directly the solar energy without transforming it. If light storage will be possible by using plasmons, or laser cavities, or light trapping as in the black body model, it will be possible too to use directly the solar energy for lighting. The optical manipulation and optical engines concepts were already demonstrated experimentally [14]. The progress in photonics with optical circuits, optical transistors, etc shows that the photonics or plasmonics informatics might be possible too. If the laser propulsion will be achieved, and the optical engines will work, means that we might have also motors working with light and nor with electricity.

In my personal point of view, one of the most important research aims is to preserve our world. Today we use solar cells panels to transform the solar energy in electricity for lighting, heating, motors, communications etc. Tomorrow maybe we will use directly the solar energy for all these ends. In this direction, the biggest challenges of tomorrow research in physics will be the plasmonics informatics and the photonics motors.